\documentclass[12pt]{article}
\usepackage{latexsym}
\usepackage[cp1250]{inputenc}

\title{Hamiltonian structures for Pais-Uhlenbeck oscillator}
\author{Katarzyna Bolonek\thanks{supported by the {\L}\'od\'z University grant No. 690.},\\ 
 Piotr  Kosi\'nski\thanks{supported by the grant 1 P03B 021 28 of the Polish Ministry of Science.} \\
Department of Theoretical Physics II \\
University of {\L}\'od\'z \\
Pomorska 149/153, 90 - 236 {\L}\'od\'z, Poland.}
\date{}

\begin{document}
\maketitle
\begin{abstract}
The Hamiltonian structures for quartic oscillator are considered. All structures admitting quadratic Hamiltonians 
are classified.
\end{abstract}

\newpage

\section{Introduction}

There is a long-held belief that in quantum theory of gravitation space-time must change its nature at distances 
comparable to the 
Planck scale. In order to model such a situation one can invoke the Heisenberg uncertainty rules.
 In their standard form they
make the notion of classical phase space meaningless on quantum level while configuration space retains its meaning. 
However, one can 
further demand that the coordinates are noncommuting operators which implies some uncertainty relations making 
also the notion
of the point in space-time no longer sensible. The simplest way to do this is to impose the commutation rules
\begin{eqnarray}
[x^{\mu},\;x^{\nu }]=i\Theta^{\mu \nu},\nonumber
\end{eqnarray}
where $\Theta^{\mu \nu}$\ is a constant $c$-number tensor. Recently, there has been much activity concerning 
field theories on such noncommutative space-time 
\cite{b1}, \cite{b2}. They appear to have some attractive properties. On the other hand their quantization seems 
to be more subtle problem than in the standard case. In fact, the noncommutative space-time can be replaced by its
 commutative counterpart provided one simultaneously replaces ordinary product of field variables by "star product"
defined by
\begin{eqnarray*}
\Phi _1(x)\star \Phi _2(x) = e^{\frac{i}{2}\Theta ^{\mu \nu }\frac{\partial }{\partial x^\mu }\frac{\partial }
{\partial y^\nu}}\Phi _1(x)\Phi _2(y)\mid _{x=y} 
\end{eqnarray*}
Therefore, once $\Theta ^{0 i }\not=0$, the resulting Lagrangian contains time derivatives of arbitrary order; 
the theory is nonlocal in time. This makes the quantization procedure much more complicated. Indeed, within the standard 
framework, the first step to quantize a given classical theory is to put it in the Hamiltonian form. There exists the
 general algorithm which allows to construct the Hamiltonian formalism for 
higher-derivative \cite{b3}, \cite{b4}, \cite{b5} and nonlocal \cite{b6}, \cite{b7}, \cite{b8} theories. However,
 its main drawback is that the Hamiltonian is not bounded from below; the quantization can be formally
 carried out but the resulting theory has serious disadvantages like, for example, the nonexistence of stable
ground (vacuum) state. This is the price one has to pay for the generality of Ostrogradski formalism. From this point of 
view it seems reasonable to pose the question whether, for a given specific system, 
there exist alternative canonical formalisms more adequate for quantization purposes. It can happen that, due to the 
peculiar properties of the system under consideration, there exists canonical formalism which, being quantized,
 produce quantum theory with more desirable properties than Ostrogradski approach. Our main motivation is to show, 
on the simplest example, that such a situation is possible; namely, that, in some cases, there exists a variety of 
Hamiltonians and the corresponding symplectic structures including those leading to the nice 
quantum theory (with stable ground state, etc.)

The Ostrogradski instability is shared by all  theories described by the Lagrangians containing time derivatives 
of at least second order. Moreover, the instability phenomenon seems to be not directly related to the nonlinear
 character of underlying dynamics. Therefore, the simplest model to be considered is the celebrated Pais-Uhlenbeck
 quartic oscillator \cite{b9}, linear theory of fourth order. 

We will study here alternative Hamiltonian formalisms for quartic oscillator.
  The starting point is the obvious observation that the 
general solution 
to Lagrangian equation depends on four arbitrary constants. This implies that the corresponding Hamiltonian system
 should have two degrees of freedom. 
By inspecting the explicit form of solutions we find that there are always at least two independent globally defined 
constants of motion which,
in addition, are quadratic in dynamical variables. On the other hand, the Hamiltonian must be also a constant of motion. 
Therefore, we can write out the most general Ansatz for quadratic 
Hamiltonian. By demanding the canonical equations to be equivalent to the initial Lagrangian one we find the relevant
 Poisson structures. In principle, the family of candidates for Hamiltonian functions is much wider. First, one could 
take an arbitrary function of two above-mentioned constants of motion. Moreover,for some values of parameters the
 dynamics is superintegrable, i.e. admits third independent globally defined constant of motion; then the most general
 Hamiltonian is a function of three integrals of motion. However, more complicated Hamiltonians result in more 
complicated or even singular symplectic structures. This implies that the relation between basic dynamical variable,
 its time derivatives and Darboux coordinates is a complicated nonlinear one and it is not clear whether it can be
promoted to quantum theory.\\ 
Let us conclude this section with the following remark. Our construction is neither a pure application nor 
 an extension of Ostrogradski algorithm. First, it cannot be applicable for all quartic systems.
 This can be seen by noting that we need here second ( independent of the Ostrogradski Hamiltonian) globally
 defined integral of motion. 
This implies that Ostrogradski dynamics is integrable which doesn't seem to be automatically true, in spite
of the fact that the canonical equations for Ostrogradski Hamiltonian have a very specific form. Second, the family 
of Hamiltonians constructed here includes in some cases the positive-definite ones,
 the property not shared by Ostrogradski Hamiltonian.\\
The details of our
construction are presented in Sec.II while Sec.III is devoted to concluding remarks.
Appendix contains some additional remarks concerning the problem of embedding the fourth-order system into Lagrangian
system of two degrees of freedom.

\section{Hamiltonian structures}

Our starting point is the following Lagrangian 
\begin{eqnarray}
L=\frac{m}{2}\dot{q}^2-\frac{m\omega^2}{2}q^2-\frac{m\lambda}{2}\ddot{q}^2\label{w1}
\end{eqnarray}
For $\lambda =0$\ one gets the harmonic oscillator of mass $m$\ and frequency $\omega$.\\
The relevant dynamical equation reads
\begin{eqnarray}
\lambda q^{(IV)}+\ddot{q}+\omega^2q=0\label{w2}
\end{eqnarray}
or, equivalently
\begin{eqnarray}
\lambda\left(\frac{d^2}{dt^2}+\omega^2_1\right)\left(\frac{d^2}{dt^2}+\omega_2^2\right)q=0;\label{w3}
\end{eqnarray}
here
\begin{eqnarray}
w^2_{1,\;2}\equiv \frac{1\pm\sqrt{1-4\lambda\omega^2}}{2\lambda}\label{w4}
\end{eqnarray}

The form of solution to eq. (\ref{w3}) depends on $w^2_{1,\;2}$. There are the following possibilities:\\
(i) $0<\;\lambda <\;\frac{1}{4\omega^2}$; then $\omega^2_{1,\;2}>0$\ and $\omega^2_1\neq \omega^2_2$;\\
(ii) $\lambda =0$; the harmonic oscillator case \\
(iii) $\lambda<0$; then $\omega^2_1<0,\;\omega^2_2>0,\; \omega^2_1+\omega_2^2\neq 0$\ \\
(iv) $\lambda =\;\frac{1}{4\omega^2}$; then $\omega_1^2=\omega^2_2=2\omega^2$, i.e. we are dealing with degeneracy\\
(v) $\lambda >\frac{1}{4\omega^2}$; both $\omega^2_{1,\;2}$\ are complex
\begin{eqnarray}
w_{1,\;2}^2=\frac{1\pm i\sqrt{4\lambda\omega^2-1}}{2\lambda},\;\;\omega_1=\bar{\omega_2}\equiv\omega_0\label{w5}
\end{eqnarray}
We shall consider these cases separately.\\

(1) \underline{The oscillatory regime ((i))}
\\

The general solution reads here
\begin{eqnarray}
q(t)=A_1cos(\omega_1t+\alpha_1)+A_2cos(\omega_2t+\alpha_2)\label{w6}
\end{eqnarray}

It depends on four arbitrary constants $A_{1,\;2},\;\alpha_{1,\;2}$\ which can be found knowing $q,\;\dot{q},\;\ddot{q}$\
 and $\stackrel{\ldots}{q}$\ 
at any given time. Consequently, there are at most four independent locally defined integrals of motion; 
however, at least one of them must depend
explicitly on time. Two integrals can be readily found by computing $A^2_{1,\;2}$\ from eq. (\ref{w6}) 
and its first three time derivatives. In this
way one obtains the global integrals (normalised for further convenience) 
\begin{eqnarray}
&&J_1=\frac{m}{\sqrt{2}(\omega^4_1-\omega^4_2)}\left((\stackrel{\ldots}{q}+\omega^2_1\dot{q})^2+\omega_2^2(\ddot{q}+\omega^2_1q)^2\right)\nonumber \\
&&J_2=\frac{m}{\sqrt{2}(\omega^4_1-\omega^4_2)}\left((\stackrel{\ldots}{q}+\omega^2_2\dot{q})^2+\omega_1^2(\ddot{q}+\omega^2_2q)^2\right)\label{w7}
\end{eqnarray}

For generic values of parameters no additional independent globally defined integral (which does not depend explicitly 
on time) exists;
our system is integrable but not superintegrable. However, for $\lambda,\;\omega^2$\ such that $\frac{\omega_1}{\omega_2}$\ is rational, 
$\frac{\omega_1}{\omega_2}=\frac{k}{l}$, it becomes superintegrable. The additional integral can be constructed
 as follows \cite{b10}. One writes
$sin(l\alpha_1-k\alpha_2)=sin(l(\omega_1t+\alpha_1)-k(\omega_2t+\alpha_2))$; the latter is expressible polynomially in 
$sin(\omega_{1,\;2}t+\alpha_{1,\;2}), \; cos(\omega_{1,\;2}t+\alpha_{1,\;2})$\ which, in turn, can be computed from eq. (\ref{w6}) and its first three
time derivatives. In what follows we are interested in generic values of $\lambda$. Therefore, we consider $J_{1,\;2}$\ to be the only relevant integrals.

As usual, the integrals of motion are related to some symmetries. Using Noether theorem suitably generalised to higher-derivative theories one finds the
symmetries responsible for the existence of $J_{1,\;2}$. They read
\begin{eqnarray}
q\rightarrow q+\varepsilon (\stackrel{\ldots}{q}\pm(\omega_1^2-\omega_2^2)\dot{q})\label{w8}
\end{eqnarray}
We can now construct the Hamiltonian formalism. There exists standard procedure called Ostrogradski formalism
 \cite{b3}, \cite{b4} which works for any 
higher-derivative theory. However, for a particular dynamics there can exist a variety of suitable Hamiltonian structures.

For the reason explained in Sec.I we restrict ourselves to quadratic Hamiltonians. Keeping in mind that the Hamiltonian itself is a constant of motion and, moreover, its rescaling
is equivalent to the time rescaling, one can write the following Ansatz
\begin{eqnarray}
H(\beta )=J_1cos\beta +J_2sin\beta ,\;\;-\pi\leq\beta\;<\pi , \label{w9}
\end{eqnarray}
Using
\begin{eqnarray}
q^{(n)}=\{q^{(n-1)},\;H\},\;\;n=1,\;2,\;3\label{w10}
\end{eqnarray}
one finds the following one-parameter family of Poisson structures
\begin{eqnarray}
&&\{q,\;\dot{q}\}=\gamma (\frac{1}{cos\beta}+\frac{1}{\sin\beta})\nonumber \\
&&\{q,\;\ddot{q}\}=0\nonumber \\
&&\{q,\;\stackrel{\ldots}{q}\}=-\gamma (\frac{\omega_2^2}{cos\beta}+\frac{\omega_1^2}{\sin\beta})\label{w11} \\
&&\{\dot{q},\;\ddot{q}\}=\gamma (\frac{\omega_2^2}{cos\beta}+\frac{\omega_1^2}{\sin\beta})\nonumber \\
&&\{\dot{q},\;\stackrel{\ldots}{q}\}=0\nonumber \\
&&\{\ddot{q},\;\stackrel{\ldots}{q}\}=\gamma (\frac{\omega^4_2}{cos\beta}+\frac{\omega^4_1}{\sin\beta})\nonumber 
\end{eqnarray}
with 

\begin{eqnarray}
\gamma \equiv \frac{1}{\sqrt{2}m\lambda (\omega_1^2-\omega_2^2)}\label{w12}
\end{eqnarray}
Let us note the following:\\
(a) The Poisson structure exists for all $\beta$\ except $\beta= -\pi, \;-\frac{\pi}{2},\;0,\;\frac{\pi}{2}$\ 
(this can be easily understood from 
eq. (\ref{w6})- both $A_1$\ $A_2$\ are needed to characterize fully the motion); consequently, there exists four disjoint
 sectors for
 $\beta : \;(-\pi,\;-\frac{\pi}{2}),\; (-\frac{\pi}{2},\;0),\; (0,\;\frac{\pi}{2})$\ and $(\frac{\pi}{2},\;\pi)$. However, the transformation
$H\rightarrow -H,\;q_i \leftrightarrow p_i$\ leaves Hamiltonian equations invariant. Therefore, it is sufficient to
 consider the sectors
$(-\frac{\pi}{2},\;0)$\ and $(0,\; \frac{\pi}{2})$.\\
(b) One easily checks that 
\begin{eqnarray}
det\left[\{q^{(m)},\;q^{(n)}\}^3_{m,\;n=0}\right]=\left(\frac{\omega^2_1-\omega_2^2}{cos\beta\; sin\beta}\right)^2\label{w13}
\end{eqnarray}
which is nonvanishing. We conclude that the Poisson structures (\ref{w11}) are sympletic.\\
(c) For any admissible $\beta$\
\begin{eqnarray}
&&\{\ddot{q}+\omega _1^2q,\; \ddot {q}+\omega _2^2q\} =0 \nonumber \\
&&\{\stackrel{\ldots}{q}+\omega _1^2\dot {q}, \; \stackrel{\ldots}{q}+\omega _2^2\dot {q}\} =0 \nonumber \\
&&\{\ddot {q}+\omega _1^2q, \; \stackrel{\ldots}{q}+\omega _2^2\dot {q}\} =0\label{w14}\\
&&\{\ddot {q}+\omega _2^2q, \; \stackrel{\ldots}{q}+\omega_ 1^2\dot {q}\}=0 \nonumber
\end{eqnarray}
(d) The structures corresponding to different $\beta 's$\ are different (i.e. not canonically equivalent).
Indeed, $q,\dot {q},\ddot {q}, and \stackrel{\ldots}{q}$\ are well-defined functions of canonical variables. 
Therefore, the canonical transformations cannot change the numerical values of the Poisson brackets. On the other hand, 
due to $\omega^2_1  \not= \omega^2 _2$, $sin\beta $ and $cos\beta $\ are uniquely fixed once the RHS 
of eq. (\ref{w11}) are known.\\
The canonical variables are found by passing to Darboux coordinates. There is a freedom in defining such a transformation 
- one can always perform an additional symplectic ( in our case - also linear ) transformation. We shall impose a further 
constraint $p_i\sim \dot {q}_i ,\; i=1,2$. Using (c) one finds the following canonical variables:\\
 - for the $ (0,\frac{\pi  }{2})$\ sector:
\begin{eqnarray}
&&q_1=\delta \sqrt{\cos \beta }(\ddot {q}+\omega _1^2q) \nonumber \\
&&p_1=m\delta \sqrt{\cos \beta }(\stackrel{\ldots}{q}+\omega _1^2\dot {q}) \nonumber \\
&& q_2=\delta \sqrt{\sin \beta }(\ddot {q}+\omega _2^2q) \nonumber \\
&& p_2=m\delta \sqrt{\sin\beta }(\stackrel{\ldots}{q}+\omega _2^2\dot {q}) \label{w15}\\
&& q=\frac{1}{\sqrt{\sqrt{2}\lambda (\omega _1^2-\omega _2^2)}}(\frac{q_1}{\sqrt{cos\beta }}-\frac{q_2}{\sqrt{sin\beta }})\nonumber\\
&& H(\beta )=(\frac{p_1^2}{2m}+\frac{m\omega _2^2}{2}q_1^2)+(\frac{p_2^2}{2m}+\frac{m\omega _1^2}{2}q_2^2)\nonumber\\
&& \delta \equiv \sqrt{\frac{\sqrt{2}\lambda }{\omega _1^2-\omega _2^2}} \nonumber
\end{eqnarray}
-for the $(-\frac{\pi }{2},0)$\ sector: 
\begin{eqnarray}
&& q_1=\delta \sqrt{cos\beta }(\ddot {q}+\omega _1^2q)\nonumber \\
&& p_1=m\delta \sqrt{cos\beta }(\stackrel{\ldots}{q}+\omega _1^2\dot {q})\nonumber \\
&& q_2=\delta \sqrt{-sin\beta }(\ddot {q}+\omega _2^2q)\nonumber \\
&& p_2=-m\delta \sqrt{-sin\beta }(\stackrel{\ldots}{q}+\omega_2^2\dot {q}) \label{w16} \\
&& q=\frac{1}{\sqrt{\sqrt{2}\lambda (\omega _1^2-\omega _2^2)}}(\frac{q_1}{\sqrt{cos\beta }}-\frac{q_2}{\sqrt{-sin\beta }})\nonumber \\
&& H(\beta )=(\frac{p_1^2}{2m}+\frac{m\omega _2^2}{2}q_1^2)-(\frac{p_2^2}{2m}+\frac{m\omega _1^2}{2}q_2^2)\nonumber
\end{eqnarray}
The formulae (15), (16) have a nice interpretation. The $q$\ variable is one of the coordinates of twodimensional quadratic 
system for which $q_1$\ and $q_2$\ are normal coordinates [11]. However, for $\beta \in (-\frac{\pi }{2},0)$\ the energy 
of one of the normal oscillations enters with negative sign. Actually, in each sector all systems look the same except the 
formula for $q$\ in terms of normal coordinates $q_{1,2}$. \\
Let us consider in some detail the sector $(-\frac{\pi }{2},0)$. Passing to the Lagrangian
\begin{eqnarray}
L(\beta )=(\frac{m}{2}\dot {q}_1^2-\frac{m\omega _2^2}{2}q_1^2)-(\frac{m}{2}\dot {q}_2^2-\frac{m\omega _1^2}{2}q_2^2)\label{w17}
\end{eqnarray}
is a regular procedure. On the other hand, under the canonical transformation
\begin{eqnarray}
&& \tilde q_1=\frac{1}{\sqrt{\lambda (\omega _1^2-\omega _2^2)}}(q_1-q_2) \nonumber \\
&& \tilde q_2=\frac{1}{m\sqrt{\lambda (\omega _1^2-\omega _2^2)}}(p_1+p_2) \label{w18}\\
&& \tilde p_1=\sqrt{\frac{\lambda }{\omega _1^2-\omega _2^2}}(\omega _1^2p_1+\omega _2^2p_2) \nonumber \\
&& \tilde p_2=m\sqrt{\frac{\lambda }{\omega _1^2-\omega _2^2}}(\omega _2^2q_1-\omega _1^2q_2) \nonumber
\end{eqnarray}
the Hamiltonian attains the Ostrogradski form
\begin{eqnarray}
H(\beta )=\tilde p_1\tilde q_2-\frac{\tilde p_2^2}{m\lambda }-\frac{m}{2}\tilde q_2^2+\frac{m\omega ^2}{2}\tilde q_1^2 \label{w19}
\end{eqnarray}
Note that eq. (19) defines now a singular Hamiltonian in the sense that momenta are not expressible in terms of velocities 
and coordinates; in fact, $\dot {\tilde q_i}=\frac{\partial H}{\partial \tilde p_i}$\ imply $\dot {\tilde q_2}=
-\frac{\tilde p_2}{m\lambda }$, but also $\dot {\tilde q_1}=\tilde q_2$, so $\tilde p_1$\ cannot be expressed in terms of 
$\tilde q_i,\;\dot {\tilde q_i}$. Therefore, some care is needed when passing to the Lagrangian formalism which results in 
additional variable - the Lagrange multiplier enforcing $\dot {\tilde q_1}=\tilde q_2$; as a result $\tilde q_1$\ obeys 
eq.(2). However, $\tilde q_1$\ coincides with $q$\ only for $\beta = - \frac{\pi }{2}$; for other $\beta 's \; q$\ and 
$\tilde q_1$, are different linear combinations of normal coordinates $q_{1,2}$.\\
 Using eqs. (16) one can express the Lagrangian (17) in terms of $q$\ variable (up to a total derivative)
\begin{eqnarray}
&& L(\beta )=\frac{m\delta ^2}{2}(cos\beta +sin\beta )\stackrel{\ldots}{q}^2 + \nonumber \\
&& -m\delta ^2(\omega_1^2(cos\beta +\frac{1}{2}sin\beta )+
\omega _2^2(sin\beta +\frac{1}{2}cos\beta ))\ddot {q}^2+ \nonumber \\
&& +\frac{m\delta ^2}{2}((\omega _1^2+2\omega _2^2)\omega_1^2cos\beta +(\omega _2^2+2\omega _1^2)\omega _2^2sin\beta )
\dot {q}^2+ \nonumber \\
&& -\frac{m\delta ^2}{2}\omega _1^2\omega_2^2(\omega_1^2cos\beta +\omega _2^2sin\beta )q^2 \label{w20}
\end{eqnarray} 
which leads to the following equation of motion
\begin{eqnarray}
((cos\beta +sin\beta )\frac{d^2}{dt^2}+(\omega _1^2cos\beta +\omega _2^2sin\beta ))(\frac{d^2}{dt^2}+\omega _1^2)
(\frac{d^2}{dt^2}+\omega _2^2)q=0 \label{w21}
\end{eqnarray}
We see from eq.(\ref{w21}) that there appears a new mode $\omega ^2 = \frac{\omega _1^2cos\beta  + \omega _2^2sin\beta }
{cos\beta  + sin\beta }$\ unless $\beta  = -\frac{\pi }{4}$. This is not surprising. First, let us stress that the theory
defined by eqs. (\ref{w16}) and (\ref{w17}) solves the problem of finding the Hamiltonian system containing eq. (\ref{w2})
( or, equivalently, eq. (\ref{w3})) as one of dynamical equations. Indeed, eq.(\ref{w2}) is the direct cosequence of
 the definition of $q$\ in terms of $q_{1,2}$\ and the basic dynamical equations the latter obey. Moreover, due to 
$\omega _1 \neq \omega _2$, in order to determine time-dependence of $q$\ one has to know both $q_1$\ and $q_2$\
which implies one has to impose four initial conditions on $q$; therefore, the theory given by eqs. (\ref{w16}) and
(\ref{w17}) describes the general solution to eq.(\ref{w2}) for arbitrary value of $\beta$. In fact, the present 
formulation does not differ very much in spirit from Ostrogradski formalism. In the latter one of the canonical 
equations implies that the substitution $q_1\rightarrow q,\;q_2\rightarrow \dot q $\ is cosistent while in the
 former Hamiltonian equations imply the consistency of the rule $q_{1,2}\sim \ddot q+\omega ^2_{1,2}q$. For this 
reason, making the substitution $q_i\sim \ddot q+\omega ^2_iq$\ in (\ref{w17}) one gets consistent equation in 
spite of the fact that the number of independent variables is reduced. On the other hand this is not a point 
transformation which in general results in new modes (see Appendix).\\    
Finally, let us compare our findings with those of Ref.[9]. Again, it is a matter of simple computation to verify 
that the formalism developed in Sec. IIA of [9] corresponds to $\beta =-\frac{\pi }{4}$. \\

(2) \underline{The case $\lambda <0$\ ((iii))}\\

Let us pass to the case (iii). Putting $\omega _1^2=-\mid \omega _1\mid ^2$\ one gets 
\begin{eqnarray}
(\frac{d^2}{dt^2}-\mid \omega _1\mid ^2)(\frac{d^2}{dt^2}+\omega _2^2)q=0 \label{w22}
\end{eqnarray}
with the general solution
\begin{eqnarray}
q=Ae^{\mid \omega _1\mid t}+A'e^{-\mid \omega _1\mid t}+Bcos(\omega _2t+\beta ) \label{w23}
\end{eqnarray}
$As$\ in the previous case one easily finds two integrals.
\begin{eqnarray}
&& I_1=\frac{m}{\sqrt{2}(\mid \omega _1\mid ^4-\omega _2^4)}((\stackrel{\ldots}{q} -\mid \omega _1\mid ^2\dot q)^2+
\omega_2^2(\ddot q-\mid 
\omega _1\mid ^2q)^2)\nonumber \\
&& I_2=\frac{m}{\sqrt{2}(\mid \omega _1\mid ^4-\omega _2^4)}((\stackrel{\ldots}{q} +\omega _2^2\dot q)^2-\mid \omega _1\mid ^2(\ddot q+
\omega _2^2q)^2) \label{w24}
\end{eqnarray}
However, there exists also the third globally defined integral. The reason for that is that there is now only one angle 
variable which has to be cyclic. The additional integral can be found by computing $ln (e^{\mid \omega _1\mid t})\equiv 
\mid \omega _1\mid t+ln A$\ and $cos(\omega _2t+\beta )$. Then $C\equiv arc cos(cos(\omega _2t+\beta ))-\frac{\omega _2}
{\mid \omega _1\mid }ln (Ae^{\mid \omega _1\mid t})$\ is time - independent and $cos C$\ can be computed from $q,\dot q,
\ddot q$\ and $\stackrel{\ldots}{q} $. The resulting expression is rather complicated and will not be considered here. \\
We proceed along the same lines as in the first case. Define the Hamiltonian
\begin{eqnarray}
H(\beta )=I_1cos\beta +I_2sin\beta , \;\;\;-\pi \leq \beta <\pi  \label{w25}
\end{eqnarray}
The family of admissible Poisson structures reads
\begin{eqnarray}
&& \{q,\dot q\}=\gamma (\frac{1}{cos\beta }+\frac{1}{sin\beta }) \nonumber \\
&& \{q,\ddot q\}=0 \nonumber \\
&& \{q,\stackrel{\ldots}{q}\}=\gamma (\frac{\mid \omega _1\mid ^2}{sin\beta }-\frac{\omega _2^2}{cos\beta }) \nonumber \\
&&  \{\dot q,\ddot q\}=-\gamma (\frac{\mid \omega _1\mid ^2}{sin\beta }-\frac{\omega _2^2}{cos\beta }) \label{w26} \\
&& \{\dot q,\stackrel{\ldots}{q}\}=0 \nonumber \\
&& \{\ddot q,\stackrel{\ldots}{q}\}=\gamma (\frac{\mid \omega _1\mid ^4}{sin\beta }+\frac{\omega _2^4}{cos\beta }) \nonumber \\
&& \gamma =\frac{1}{-\sqrt{2}m\lambda (\mid \omega _1\mid ^2+\omega _2^2)} \nonumber
\end{eqnarray}
These structures can be obtained from eq.(11) making the replacement $\omega _1^2\rightarrow -\mid \omega _1\mid ^2$. 
Again we have four sectors $\beta $\ and it is sufficient to consider two of them only:\\
- for $\beta \in (0,\frac{\pi }{2})$\ one gets 
\begin{eqnarray}
&& q_1=\delta \sqrt{cos\beta }(\ddot q-\mid \omega _1\mid ^2q) \nonumber \\
&& p_1=m\delta \sqrt{cos\beta }(\stackrel{\ldots}{q}-\mid \omega _1\mid ^2\dot q) \nonumber \\
&&q_2=\delta \sqrt{sin\beta }(\ddot q+\omega _2^2q) \nonumber \\
&& p_2=m\delta \sqrt{sin\beta }(\stackrel{\ldots}{q}+\omega _2^2\dot q) \label{w27} \\
&& q=\frac{1}{\sqrt{-\sqrt{2}\lambda (\mid \omega _1\mid ^2+\omega _2^2)}}(\frac{q_1}{\sqrt{cos\beta }}-\frac{q_2}{\sqrt
{sin\beta }}) \nonumber \\
&& H(\beta )=(\frac{p_1^2}{2m}+\frac{m\omega _2^2}{2}q_1^2)+(\frac{p_2^2}{2m}-\frac{m\mid \omega _1\mid ^2}{2}q_2^2)\nonumber\\
&& \delta \equiv \sqrt{\frac{-\sqrt{2}\lambda }{\mid \omega _1^2\mid +\omega _2^2}} \nonumber
\end{eqnarray}
- for $\beta \in (-\frac{\pi }{2},0)$\ one gets 
\begin{eqnarray}
&& q_1=\delta \sqrt{cos\beta }(\ddot q-\mid \omega _1\mid ^2q) \nonumber \\
&& p_1=m\delta \sqrt{cos\beta }(\stackrel{\ldots}{q}-\mid \omega _1\mid ^2\dot q) \nonumber \\
&&q_2=\delta \sqrt{-sin\beta }(\ddot q+\omega _2^2q) \nonumber \\
&& p_2=-m\delta \sqrt{-sin\beta }(\stackrel{\ldots}{q}+\omega _2^2\dot q) \label{w28} \\
&& q=\frac{1}{\sqrt{-\sqrt{2}\lambda (\mid \omega _1\mid ^2+\omega _2^2)}}(\frac{q_1}{\sqrt{cos\beta }}-\frac{q_2}{\sqrt
{-sin\beta }}) \nonumber \\
&& H(\beta )=(\frac{p_1^2}{2m}+\frac{m\omega _2^2}{2}q_1^2)-(\frac{p_2^2}{2m}-\frac{m\mid \omega _1\mid ^2}{2}q_2^2) \nonumber
\end{eqnarray}

Again the conclusion is that the $q$\ variable is a linear combination of normal coordinates for some quadratic system. 
The only difference as compared with the previous case is that the forces are in part repelling.\\

(3) \underline{The degenerate case} ((iv))\\

Consider the double frequency case (iv):
\begin{eqnarray}
(\frac{d^2}{dt^2}+2\omega ^2)^2q=0 \label{w29}
\end{eqnarray}
Then
\begin{eqnarray}
q(t)=A_1cos(\sqrt{2}\omega t+\alpha_1)+A_2tcos(\sqrt{2}\omega t+\alpha _2) \label{w30}
\end{eqnarray}
The relevant integrals of motion are (again suitably normalised)
\begin{eqnarray}
&& I_1=\frac{m}{\omega ^4}((\stackrel{\ldots}{q} +2\omega ^2\dot q)^2+2\omega ^2(\ddot q+2\omega ^2q)^2) \nonumber \\
&& I_2=\frac{m}{\omega ^2}(2(\stackrel{\ldots}{q}+2\omega ^2\dot q)\dot q-\ddot {q}^2+4\omega ^4q^2) \label{w31}
\end{eqnarray}
Again, our system admits third integral which is globally defined but complicated and won't be considered.\\
We put
\begin{eqnarray}
H(\beta )=I_1cos\beta +I_2sin\beta \;\;\; -\pi \leq \beta <\pi  \label{w32}
\end{eqnarray}
and find
\begin{eqnarray}
&& \{q,\dot q\}=-\frac{cos\beta }{2msin^2\beta } \nonumber \\
&& \{q,\ddot q\}=0 \nonumber \\
&& \{q,\stackrel{\ldots}{q}\}=\frac{(2cos\beta +sin\beta )\omega^2}{2msin^2\beta } \nonumber \\
&& \{\dot q,\ddot q\}=-\frac{(2cos\beta +sin\beta )\omega ^2}{2msin^2\beta } \label{w33} \\
&& \{\dot q,\stackrel{\ldots}{q}\}=0 \nonumber\\
&& \{\ddot q,\stackrel{\ldots}{q}\}=-\frac{2(cos\beta +sin\beta )\omega ^4}{msin^2\beta } \nonumber
\end{eqnarray}
There are now two sectors, $(-\pi ,0)$\ and $(0,\pi )$\
and its sufficient to consider only one, say $\beta \in (0,\;\pi )$. Note that for $\beta =\frac{\pi}{2},\;H=J_2$;
this is possible because $J_2$\ depends on both $A_1$\ and $A_2$.

Let us define new variables
\begin{eqnarray}
&&q_1=\frac{\sqrt{sin\beta}}{\omega^2}(\ddot{q}+2\omega^2q)\nonumber \\
&&q_2=\frac{cos\beta}{\sqrt{sin\beta}\omega^2}(\ddot{q}+2(1+tg\beta )\omega^2q) \label{w34} \\
&&p_1=\frac{mcos\beta}{\sqrt{sin\beta}\omega^2}(\stackrel{\ldots}{q}+2(1+tg\beta )\omega^2\dot{q})\nonumber \\
&&p_2=\frac{m\sqrt{sin\beta}}{\omega^2}(\stackrel{\ldots}{q}+2\omega^2\dot{q})\nonumber
\end{eqnarray}
Then the Hamiltonian takes the form
\begin{eqnarray}
H=\frac{p_1p_2}{m}+m\omega^2(2q_1q_2-q_1^2)\label{w35}
\end{eqnarray}
while $q$\ is the linear combination of basic variables
\begin{eqnarray}
q=\frac{cos\beta}{2\sqrt{sin\beta}}\left(-\frac{q_1}{sin\beta}+\frac{q_2}{cos\beta}\right)\label{w36}
\end{eqnarray}

The Hamiltonian does not depend explicitly on $\beta$\ and the only $\beta$-dependence comes from the
 expression for $q$\ in terms of basic variables $q_1,\;q_2$.\\

\underline{Complex frequencies ((v))}\\

Finally, let us consider the complex frequencies case. Formally, one can use the results of (i) and define the integrals
\begin{eqnarray}
&&J_1=\frac{m}{\sqrt{2}(\omega^4_0-\bar{\omega}^4_0)}((\stackrel{\ldots}{q}+\omega^2_0\dot{q})^2+\bar{\omega}^2_0
(\ddot{q}+\omega_0^2q)^2)\label{w37} \\
&&J_2=\frac{m}{\sqrt{2}(\omega^4_0-\bar{\omega}^4_0)}((\stackrel{\ldots}{q}+\bar{\omega}^2_0\dot{q})^2+\omega^2_0
(\ddot{q}+\bar{\omega}_0^2q)^2)\nonumber
\end{eqnarray}
They are no longer real but rather obey 
\begin{eqnarray}
\bar{J}_2=-J_1\label{w38}
\end{eqnarray}

The one-parameter Ansatz for the real Hamiltonian reads
\begin{eqnarray}
H(\beta )=i(e^{i\beta}J_1+e^{-i\beta}J_2)\label{w39}
\end{eqnarray}

The relevant Poisson structure is given by
\begin{eqnarray}
&&\{q,\;\dot{q}\}=-i\gamma (e^{i\beta}+e^{-i\beta})\nonumber \\
&&\{q,\;\ddot{q}\}=0\nonumber \\
&&\{q,\;\stackrel{\ldots}{q}\}=i\gamma(\omega^2_0e^{i\beta}+\bar{\omega}_0^2e^{-i\beta})\label{w40}\\
&&\{\dot{q},\;\ddot{q}\}=-i\gamma(e^{i\beta}\omega^2_0+e^{-i\beta}\bar{\omega}^2_0)\nonumber \\
&&\{\dot{q},\;\stackrel{\ldots}{q}\}=0\nonumber \\
&&\{\ddot{q},\;\stackrel{\ldots}{q}\}=-i\gamma(\omega^4_0e^{i\beta}+\bar{\omega}^4_0e^{-i\beta})\nonumber \\
&&\gamma \equiv \frac{1}{\sqrt{2}m\lambda (\omega^2_0-\bar{\omega}_0^2)}\nonumber
\end{eqnarray}

Now, all values of $-\pi \leq \beta <\pi$\ are admissible. Again, we could consider only half of this domain, say $0\leq \beta <\pi$, but 
there is no point to do this as we are dealing with one sector only. Define
\begin{eqnarray}
&&q_1=\frac{1}{\varepsilon}(\ddot{q}+\omega_0^2q)\nonumber \\
&&q_2=\frac{1}{\bar{\varepsilon}}(\ddot{q}+\bar{\omega}_0^2)\nonumber \\
&&p_1=\frac{m}{\varepsilon}(\stackrel{\ldots}{q}+\omega_0^2\dot{q})\label{w41} \\
&&p_2=\frac{m}{\bar{\varepsilon}}(\stackrel{\ldots}{q}+\bar{\omega}_0^2\dot{q})\nonumber \\
&&\varepsilon^2\equiv -im\gamma (\omega^2_0-\bar{\omega}_0^2)^2e^{-i\beta}\nonumber
\end{eqnarray}
The Hamiltonian takes the form
\begin{eqnarray}
H(\beta)=\left(\frac{p_1^2}{2m}+\frac{m\bar{\omega}_0^2}{2}q_1^2\right)+\left(\frac{p_2^2}{2m}+\frac{m\omega^2_0}{2}q_2^2\right)\label{w42}
\end{eqnarray}
while the expression for q reads
\begin{eqnarray}
q=\frac{\varepsilon q_1-\bar{\varepsilon}q_2}{\omega_0^2-\bar{\omega}_0^2}\label{w43}
\end{eqnarray}

The canonical variables are not real. In fact, $\bar{q}_1=q_2,\;\bar{p_1}=p_2$. The real canonical variables are obtained by taking the real and imaginary parts
\begin{eqnarray}
&&q_1=\frac{1}{\sqrt{2}}(Q_1+iQ_2)\nonumber \\
&&q_2=\frac{1}{\sqrt{2}}(Q_1-iQ_2)\label{w44} \\
&&p_1=\frac{1}{\sqrt{2}}(P_1-iP_2) \nonumber \\
&&p_2=\frac{1}{\sqrt{2}}(P_1+iP_2)\nonumber
\end{eqnarray}

Then
\begin{eqnarray}
H(\beta )=\left(\frac{P_1^2}{2m}+\frac{m(\omega_0^2+\bar{\omega}_0^2)}{4}Q_1^2\right)-\left(
\frac{P_2^2}{2m}+\frac{m(\omega^2_0+\bar{\omega}_0^2)}{4}Q_2^2\right)+\frac{im}{2}(\bar{\omega}_0^2-\omega_0^2)
Q_1Q_2\label{w45}
\end{eqnarray}

Further change of variables transforming $H(\beta )$\ into the sum of dilatation and rotation is also possible
 \cite{b9} (cf. Sec. III).

\section{Concluding remarks}

Let us summarize our results. We have found essentially one-parameter families of inequivalent quadratic 
Hamiltonian structures in all cases (i), (iii) $\div$\ (v). In the first two cases these families consist of four disjoint sectors
 while there are only two sectors in the (iv) case and one - in the (v) case. In each sector the Hamiltonian can be put into the parameter-independent form; the structures belonging to any sector differ in the way the $q$-variable is expressed in terms of basic variables. 

Due to the symmetry $H\leftrightarrow -H,\;q_i \leftrightarrow p_i$, one can reduce by two the number of sectors
we have to consider. Therefore, in the case (i) one has basically two sectors. The Hamiltonian is, respectively, 
the sum or
difference of two independent harmonic oscillators. Our $q$ variable is a linear combination of two basic coordinates 
$q_1,\;q_2$. Taking into account the possibility of rescaling the Hamiltonian and performing simple canonical 
transformation $q_1\rightarrow \pm q_i,\;p_i\rightarrow \pm p_i$\ one concludes from eq. (\ref{w15}), (\ref{w16}) that $q$\
can be arbitrary linear combination of $q_1,\;q_2$\ except that both coefficients are nonvanishing.

Similar results hold for the case (iii). The only difference is that now one oscillator describes the repelling linear force. 
In the degenerate case (iv) there is essentially one sector (if one again takes into account the symmetry 
$q_i\leftrightarrow p_i,\;H\leftrightarrow -H$). The Hamiltonian takes less familiar form (\ref{w35}) while $q$\ is given
by (\ref{w36}).

In the complex case (v) there is one sector even without taking into account the above-mentioned symmetry. The Hamiltonian 
(\ref{w45}) is now the difference of two harmonic oscillators coupled by the interaction term proportional to the
product of coordinate variables. Due to the fact that the kinetic energy is not  positive definite passing
to normal coordinates is now impossible.

Obviously, the Hamiltonian structures considered in Ref. \cite{b9} are the particular elements of our families
($\beta =-\frac{\pi}{4}$\ for (i) and (iii), $\beta=\frac{\pi}{2}$\ for (iv) and $\beta=-\pi$\ for (v)).

Finally, note that, apart from the Hamiltonian, there is always an additional quadratic integral of motion. Therefore,
we expect in all cases the separation of variables is possible. This is obvious for the first two families. 
In the degenerate case one defines \cite{b9}
\begin{eqnarray}
&&q_1=q'_1\nonumber \\
&&q_2=q_1'-\frac{1}{m\omega}p'_2\label{w46} \\
&&p_1=p_1'-m\omega q_2'\nonumber \\
&&p_2=m\omega q_2'\nonumber
\end{eqnarray}
Then $H$, eq. (\ref{w35}) takes the form
\begin{eqnarray}
H=-\omega (q'_1p'_2-q_2'p_1')-m\omega^2(q_1'^2+q_2'^2) \label{w47}
\end{eqnarray}
which separates in polar coordinates. 

Finally, consider the case of two complex conjugated frequencies squared. Making an Ansatz \cite{b9}
\begin{eqnarray}
&&q_1=\frac{1}{2\sqrt{\bar{\omega_0}}}((\tilde{q}_1-\tilde{p}_2)-i(\tilde{p}_1-\tilde{q}_2)) \nonumber \\
&&q_2=\frac{1}{2\sqrt{\omega_0}}((\tilde{q}_1-\tilde{p}_2)+i(\tilde{p}_1-\tilde{q}_2)) \nonumber \\
&&p_1=\frac{1}{2}\sqrt{\bar{\omega_0}}((\tilde{q}_2+\tilde{p}_1)-i(\tilde{p}_2+\tilde{q}_1))\label{w48} \\
&&p_2=\frac{1}{2}\sqrt{\omega_0}((\tilde{q}_2+\tilde{p}_1)+i(\tilde{p}_2+\tilde{q}_1)) \nonumber
\end{eqnarray}
one obtains
\begin{eqnarray}
H=-(\frac{\omega_0+\bar{\omega}_0}{2})(\tilde{q}_1\tilde{p}_2-\tilde{q}_2\tilde{p}_1)+\frac{i}{2}(\omega_0-\bar{\omega}_0)(
\tilde{q}_1\tilde{p}_1+\tilde{q}_2\tilde{p}_2)\label{w49}
\end{eqnarray}
i.e. the Hamiltonian becomes a commuting sum of angular momentum and dilatation and separates in polar coordinates.

The Hamiltonian formalism provides the first step toward quantization. The standard approach based on Ostrogradski
formalism and Dirac procedure \cite{b12},\cite{b13} provides a consistent quantum theory. However, its serious drawback 
is that the quantum Hamiltonian is unbounded from below. One is not surprised that the Hamiltonian is unbounded from below 
if the classical motion is unbounded (the cases $(iii)\div (v)$\ above). On the contrary, in the case (i) the motion 
\underline{is} bounded while the Ostrogradski Hamiltonian is again unbounded. We have shown that in this case there exists 
the whole family of Hamiltonians which, after quantization, yield stable ground state.\\
We have to stress that in all cases under consideration the quantization procedure is quite simple because the Hamiltonians 
are built with the help of operators well-known from ordinary quantum mechanics: oscillator Hamiltonian, angular momentum, 
dilatation operator etc.
\section{Appendix.}

Let us discuss in more detail the problem of embedding the fourth-order dynamical system into Lagrangian system with two 
degrees of freedom. First let us note the following. Assume we have the first-order Lagrangian
\begin{eqnarray}
L=L(q,\dot q) \label{wA1}
\end{eqnarray}
Let us make the following substitution
\begin{eqnarray}
&& q=q(x,\dot x,\ddot x ) \label{wA2} \\
&& \tilde L(x,\dot x,\ddot x,\stackrel{\dots}{x})=L(q(x,\dot x,\ddot x),\dot q(x,\dot x,\ddot x,\stackrel{\dots}{x})) \nonumber
\end{eqnarray}
Then one easily derives the following identity
\begin{eqnarray}
&& \frac{\partial \tilde L}{\partial x}-\frac{d}{dt}\left(\frac{\partial \tilde L}{\partial \dot x}\right)+
\frac{d^2}{dt^2}\left(\frac
{\partial \tilde L}{\partial \ddot x}\right)-\frac{d^3}{dt^3}\left(\frac{\partial \tilde L}{\partial \stackrel{\dots}{x}}\right)= \nonumber \\
&& =\left(\frac{\partial q}{\partial x}-\frac{d}{dt}\frac{\partial q}{\partial \dot x}+\frac{d^2}{dt^2}\frac{\partial q}{\partial \ddot x}\right)
\left(\frac{\partial L}{\partial q}-\frac{d}{dt}(\frac{\partial L}{\partial \dot q})\right) \label{wA3}
\end{eqnarray}
We see that, in general, the new equation of motion contains additional solutions except the case $\frac{\partial q}
{\partial \dot x}=0,\; \frac{\partial q}{\partial \ddot x}=0$; in the latter case (\ref{wA2}) describes point transformation 
leading to the equivalent dynamics.\\
Consider now the system of two decoupled degrees of freedom,
\begin{eqnarray}
L=L_1(q_1,\dot q_1)+L_2(q_2,\dot q_2) \label{wA4}
\end{eqnarray}
Assume that
\begin{eqnarray}
q_i=q_i(q,\dot q,\ddot q),\;\;\; i=1,2 \label{wA5}
\end{eqnarray}
be the substitution, in terms of one variable, consistent with the equations of motion. By the latter we mean that 
substituting (\ref{wA5}) into both equations
\begin{eqnarray}
\frac{\partial L_i}{\partial q_i}-\frac{d}{dt}\left(\frac{\partial L_i}{\partial \dot q_i}\right)=0, \;\;\; i=1,2 \label{wA6}
\end{eqnarray}
produce the same equation for $q$:
\begin{eqnarray}
\frac{\partial L_i}{\partial q_i}-\frac{d}{dt}\left(\frac{\partial L_i}{\partial \dot q_i}\right)=\alpha _iF(q,\dot q,\ddot q,\stackrel
{\dots}{q},q^{(IV)}) \label{wA7}
\end{eqnarray}
with some constans $\alpha _{1,2}$. Then (\ref{wA3}) implies for the Lagrangian
\begin{eqnarray}
\tilde L(q,\dot q,\ddot q,\stackrel{\dots}{q})=L_1(q_1,\dot q_1)+L_2(q_2,\dot q_2) \label{wA7}
\end{eqnarray}
the following identity
\begin{eqnarray}
&&\frac{\partial \tilde L}{\partial q}-\frac{d}{dt}\left(\frac{\partial \tilde L}{\partial \dot q}\right)
+\frac{d^2}{dt^2}\left(\frac{\partial 
\tilde L}{\partial \ddot q}\right)-\frac{d^3}{dt^3}\left(\frac{\partial \tilde L}{\partial \stackrel{\dots}{q}}\right)
= \nonumber \\
&& =\sum\limits_{i=1}^2\left(\alpha _i\frac{\partial q_i}{\partial q}-\alpha _i\frac{d}{dt}\frac{\partial q_i}{\partial \dot q}
+\alpha _i\frac{d^2}{dt^2}\frac{\partial q_i}{\partial \ddot q}\right)F(q,\dot q,\ddot q,\stackrel{\dots}{q},q^{(IV)}) \label{wA8}
\end{eqnarray}
If it happens that $\alpha _i's$\ are such that the second and third term on RHS of (\ref{wA8}) vanish, $\tilde L$\ gives 
no additional solutions.\\
In our case
\begin{eqnarray}
L=\alpha _1\left(\frac{m\dot q_1^2}{2}-\frac{m\omega _1^2q_1^2}{2}\right)+\alpha _2\left(\frac{m\dot q_2^2}{2}-
\frac{m\omega _2^2q_2^2}{2}\right) \label{wA9}
\end{eqnarray}
and the consistent substitution reads
\begin{eqnarray*}
&& q_1\sim \ddot q+\omega _2^2q\\
&& q_2\sim \ddot q+\omega _1^2q
\end{eqnarray*}
Then $q$\ can be expressed in terms of $q_1$\ and $q_2$\ and one obtains the consistent embedding of $q$\ into twodimensional 
system of first order. Moreover, for $\alpha _1 = -\alpha _2$\ $\tilde L$\ gives no additional mode. 
 However, the procedure is consistent for \underline{any} $\alpha _{1,2}$\ provided $\alpha _1\cdot 
\alpha _2\neq 0$\ (we must have \underline{two} degrees of freedom in order to be able to express $q$\ \underline
{algebraically} in terms of basic dynamical variables).


\begin{thebibliography}{99}
\bibitem{b1}
M.R. Douglas, N. A. Nekrasov, Rev. Mod. Phys. \underline{73}, (2001),977
\bibitem{b2}
R.J. Szabo, Phys. Rep. \underline{378}, (2003), 207
\bibitem{b3}
M. Ostrogradski, Mem. Ac. St. Petrsburg  \underline{4}, (1850), 385
\bibitem{b4}
E. T. Whittaher, Analytical Dynamics, Cambridge University Press 1937
\bibitem{b5}
J. Govaerts, M. S. Rashid, hep-th/9403009
\bibitem{b6}
J. Llosa, J. Vives, J. Math. Phys. \underline{35}, (1994), 2856
\bibitem{b7}
J. Gomis, K. Kamimura, J. Llosa, Phys. Rev. \underline{D63}, (2001), 045003
\bibitem{b8}
J. Gomis, K. Kamimura, J. Ramirez, Nucl. Phys. \underline{B696}, (2004), 263
\bibitem{b9}
A. Pais, G. E. Uhlenbeck, Phys. Rev. \underline{79} (1950), 145
\bibitem{b10}
L. Landau, E. Lifshic, Mechanics, PWN, Warsaw 1966, (in Polish)
\bibitem{b11}
Tai-Chung Cheng, Pei-Ming Ho, Mao-Chuang Yeh, Nucl. Phys. \underline{B625}, (2002), 151
\bibitem{b12}
M. Henneaux, C. Teitelboim, "Quantization of Gauge Systems" ,Princeton University Press (1992)
\bibitem{b13}
P. Mannheim, A. Davidson, hep-th/0408104
\end{thebibliography}
\end{document}